# An Empirical Study of Spam and Spam Vulnerable email Accounts

Cynthia Dhinakaran and Jae Kwang Lee
*Department of Computer Engineering*
*Hannam University, South Korea*

Dhinaharan Nagamalai
*Wireilla Net Solutions Inc,*
*Chennai, India,*

## Abstract

*Spam messages muddle up users inbox, consume network resources, and build up DDoS attacks, spread malware. Our goal is to present a definite figure about the characteristics of spam and spam vulnerable email accounts. These evaluations help us to enhance the existing technology to combat spam effectively. We collected 400 thousand spam mails from a spam trap set up in a corporate mail server for a period of 14 months form January 2006 to February 2007. Spammers use common techniques to spam end users regardless of corporate server and public mail server. So we believe that our spam collection is a sample of world wide spam traffic. Studying the characteristics of this sample helps us to better understand the features of spam and spam vulnerable e-mail accounts. We believe that this analysis is highly useful to develop more efficient anti spam techniques. In our analysis we classified spam based on attachment and contents. According to our study the four years old heavy users email accounts attract more spam than four years old light users mail accounts. The 14 months old relatively new email accounts don't receive spam. In some special cases like DDoS attacks, the new email accounts receive spam. During DDoS attack 14 months old heavy users email accounts have attracted more number of spam than 14 months old light users mail accounts.*

## 1. Introduction

Email provides low cost message delivery to large number of people by simply clicking the send button. The byproducts of email like instant messaging, chat etc., make life easier and adds more sophisticated facilities to the Internet users. These days spam emerges as a serious threat to the Internet Community [8]. Spam is defined as unsolicited, unwanted mail that endangers the very existence of the e-mail system with massive and uncontrollable amounts of message [4]. Spam brings worms, viruses and unwanted data to the user's mailbox. Spammers are well organized business people or organizations that want to make money. DDoS attacks, spy ware installations, worms are not negligible portion of spam traffic. Nearly 80% of all spam are received from botnets [5]. Our aim is to present clear characteristics of spam and spam vulnerable e-mail accounts. We setup a spam trap in our mail server and collected spam for the past 14 months from January 2006 to February 2007.

We used this data for our study to characterize spam and spam vulnerable email accounts. We conducted several standard spam tests to separate spam from incoming mail traffic. The standard test includes various source filters, content filter. The various source filter tests includes Baysean filter, DNSBL, SURBL, SPF, Grey List, rDNS etc. The learning is enabled in content filters. The size of the dictionary is 50000 words. At our organization we strictly implement mail policies to avoid spam mails. The users are well instructed on how to use mail service for effective communication. We classified spam into two type likely spam mails with attachment and spam without attachment. The spam without attachment are classified as spam containing only text message and spam containing text message with URL or a clickable link. The spam with attachments are classified into four categories as spam containing Image and text, spam containing image, URL and text, spam containing Image and URL, spam containing .exe file as an attachment. The later classification is based on its contents. The .gif file attached with message in first three categories.

The rest of the paper is organized as follows. Section 2 discusses related work. Section 3 provides data collection of legitimate and spam mails. In section 4, we describe our empirical study of spam traffic. Section 5 provides details of Spam vulnerable email accounts. We conclude in section 6.

## 2. Related work

In [1] propose a novel approach to defend DDoS attack caused by spam mails. Their study reveals the effectiveness of SURBL, DNSBLs, content filters. They have presented inclusive characteristics of virus, worms and trojans accompanied spam as an attachment. Their approach is a combination of fine tuning of source filters, content filters, strictly implementing mail policies, educating user, network monitoring and logical solutions to the ongoing attack. In [8] Gomez, Crsitino presented an extensive study on characteristics of spam traffic in terms of email arrival process, size distribution, the distributions of popularity and temporal locality of email recipients etc., compared with legitimate mail traffic. Their study reveals major differences between spam and non spam mails. In [18] Cynthia et al studied extensively the characteristics of spam and spammers. Spammers use common techniques to spam end users regardless of corporate server and public mail server. In their analysis they classified spam based on attachment and contents. They observed that the spammers use software tool to send spam with attachment. The software provides sophisticated facilities to spam end users. The characteristics of spam software are hiding the sender's identity, randomly selecting text messages, identifying open relay machines, mass mailing capability, defining the spamming time and duration. They also identified that spammers manually send spam without attachment to limited number of end users. Spammers don't use spam software to send spam without attachment.

## 3. Data Collection

Our characterization of spam is based on 14 months collection of data over 400,000 spam from a corporate mail server. The web server provides service to 200 users with 20 group email IDs and 200 individual mail accounts. The speed of the Internet connection is 100 Mpbs for the LAN, with 20 Mbps upload and download speed (Due to security and privacy concerns we are not able to disclose the real domain name). To segregate spam from legitimate mail, we conducted a standard spam detection tests in our server. The spam mails detected by these techniques were directed to the spam trap in the mail server.

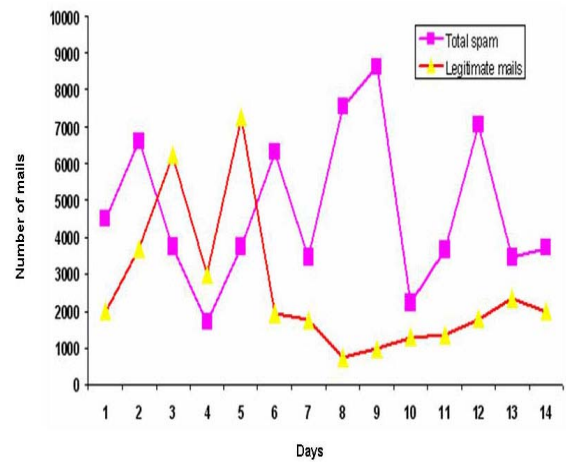

**Figure 1.** Mail traffic

The figure 1 shows the incoming mail traffic of our mail server for 2 weeks. The figure shows that the spam traffic is not related to legitimate mail traffic. The legitimate mail traffic is two way traffic induced by social network [8]. But the spam traffic is one way traffic. From this picture we can understand that the server is handling more number of spam than legitimate mail traffic. The Figure.1 shows the number of legitimate mails, and spam for a period of 2 weeks from February 1 to February 15. The x axis is day and y axis is the number of mails and spam received by the spam trap on server. Roughly the number of legitimate mails ranges from 720 to 7253 with an average rate of 906 per day. The spam mails ranges from 1701 to 8615 with an average rate of 4736 per day.

## 4. Empirical Study

We have analyzed millions of spam received in our spam trap. Mostly the spam mails are related to finance, pharmacy, business promotion, adultery services and viruses. Considerable amount of spam has virus and worms as an attachment [1]. Based on our study the spam mails typically fall into one of two camps, like spam without attachment and spam with attachment. The spam with attachment does not have any relationship with spam without attachment in terms of traffic volume [18]. Both are driven by different spammers with different technology. In our spam collection more than 50% of the spam falls under the category of spam with attachments. This kind of spam contains a combination of pictures, text and URL or a clickable link to a web site. Mostly the attachments are image file (.gif, .jpg) and executable (.exe) files. The image files are mostly .gif format and rarely .jpg format. The size of this kind of spam ranges from 5kb to 45 kb. Spam with .exe file as an attachment is mail bombs containing viruses and worms [1].

Spam with attachments can be classified into Spam mails containing image file (.gif) and text message, spam containing image, text, URL, and Spam containing only image with clickable web link, Spam containing worms, virus, Trojans as an attachments. We have monitored spam with attachment traffic for a period of 14 months from Jan 2006 to February 2007.

Spam contains text, image, URL[18]: This category of spam contains text, Image and URL. Considerable portion of these kinds of spam are responsible for phishing attacks. Spam containing Image and text: In this category we discuss spam containing text and Image as a message [18]. The text message size varies from a single line to several paragraphs. By changing the text contents and its size, the spammers try to confuse the filters. In this category we found that the majority of spam are related to pharmacy and finance. Spam containing image and URL or a clickable link: Considerable number of spam contains image file and URL or a clickable link. Since the URL or clickable link is placed inside the image, the spam can easily bypass content filters [18].

The fourth category of spam contains .exe file as an attachment. The spam with .exe file as an attachments are mostly virus, worms, trojon etc,. The attachment sizes ranges from 35 k to 140 kb. These mails are intended to spread virus, try to establish mail bombs to mount DDoS attack to the server and the network. Upon execution of the attachment, it will drop new files in windows folder and change the registry file, link to the attacker's website to download big programs to harm the network further. The infected machine collects email addresses from windows address book and automatically sends mails to others in the same domain [1].

Distributed Denial of Service (DDoS) attack is a large scale, coordinated attack on the availability of services at a victim system or network resource [5]. DDOS attack through spam mail is one of the new versions of common DDoS attack. In this type, the attacker penetrates the network by a small program attached to the spam mail. After the execution of the attached file, the mail server resources will be eaten up by mass mails from other machines in the domain results denial of services. The spam contains small size of .exe file as an attachment (for example update.exe). The attackers used double file extension to confuse the filter (Update_KB2546_*86.BAK.exe (140k)) and user. The attachment size ranges from 35 to 180 KB. The names of the worms used in these kind of DDoS attacks are WORM_start.Bt, WORM_STRAT.BG, WORM_STRAT.BR, TROJ_PDROPPER.Q. Upon execution, these worms dropped files namely serv.exe, serv.dll, serv.s, serv.wax, E1.dll, rasaw32t.dll etc. DDoS malware cause direct and indirect damage by flooding specific targets [16]. Mass mailers and network worms cause indirect damage when they clog mail servers and network bandwidth. In Network, It will consume the network bandwidth and resources, causing slow mail delivery further resulting Denial of service. The server will be down due to enormous request from clients and bulk mail processing [1].

We have analyzed the spam trap to measure the traffic of spam with virus, worm and trojon. The six week long traffic of spam with virus, worms, trojon is shown in figure 2, 3 and table 1.

**Table 1**: Spam with virus, worms and trojons traffic for 7 weeks and 7 days of first week

| Weeks | Number of Spam with virus, worms, trojons | Days of first week | Number of Spam with virus, worms, trojons |
|---|---|---|---|
| 1 | 2847 | 1 | 407 |
| 2 | 1947 | 2 | 485 |
| 3 | 1152 | 3 | 339 |
| 4 | 1834 | 4 | 253 |
| 5 | 2423 | 5 | 384 |
| 6 | 1245 | 6 | 494 |
| 7 | 2256 | 7 | 486 |

The Fig. 2 shows the number of spam containing virus, worms and trojon as an attachment for a period of seven weeks. The x axis is week and y axis is the number of spam with virus, worms and trojon as an attachment, received in our server spam trap. These types of spam received ranges from 1245 to 2847 with an average of 1957 per week. Further we analyzed these spam for a week to present single day traffic. The figure 3 shows the number of spam with virus received in our mail server for seven days. The spam mails with virus, worms and trojons ranges from 253 to 496 with an average rate of 407. The x axis is day and y axis is the number of spam with virus, worms and trojon as an attachment, received in our server spam trap. There is no relation between spam with virus traffic and legitimate mail traffic.

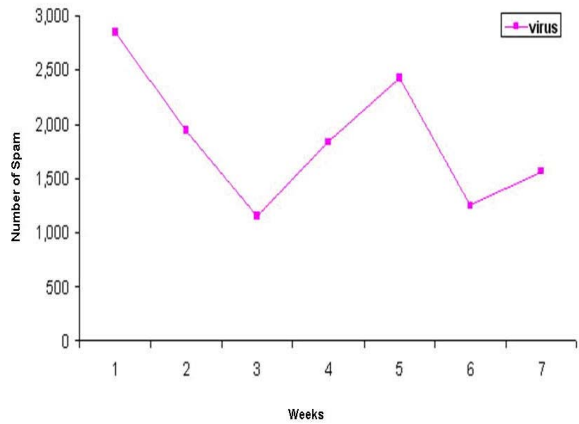

**Figure .2**. Spam with virus, worms and Trojons traffic for 7 weeks

The week long traffic of such kind of spam traffic is shown in figure 3 and table 1.

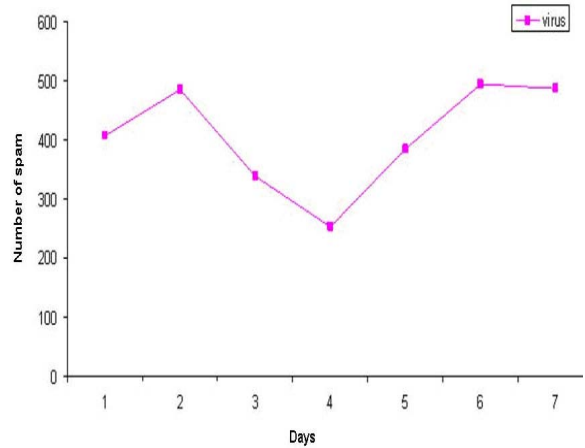

**Figure 3.** Spam with virus, worms and Trojons traffic in the first week of figure 2

From our analysis we have identified a list of frequently spammed virus, worms and trojons for a period of a month from February 1 to February 28. The list is shown in table 2.

**Table 2**: Top 10 Virus, worms, trojons received by spam tarp on February 2007

| Virus,worms,trojon | Number of spam mails |
|---|---|
| Email-Worm.Win32.NetSky.q | 630 |
| Spam.Phish.url | 3278 |
| Email-Worm.Win32.Bagle.gt | 556 |
| Spam.Porn.PORN_NB_PORNHINT_1 | 231 |
| NBH-BHIDIFM | 118 |
| Net-Worm.Win32.Mytob.c | 150 |
| Email-Worm.Win32.Zhelatin.o | 110 |
| Email-Worm.Win32.Zhelatin.u | 58 |
| Email-Worm.Win32.Bagle.gen | 97 |
| Trojan-Downloader.Win32.Agent.bet | 161 |
| Trojan-Downloader.Win32.Small.dam | 85 |

## 5. Spam vulnerable email accounts

There are many ways to get end users e-mail accounts. Many professional spammers selling e-mail accounts to spammers for cheap prices. The professional email account sellers collect e-mail accounts by using well designed software and also offer free bees. We are not going to discuss this, instead we are going to discuss about email accounts which attract more spam.

Heavy users receive more spam than others [6]. Heavy users can be defined as, those who have own web site, blogs and those actively involving in forums and chat rooms. Apart from spammers, legitimate business houses like Internet, telecom service providers also send spam to unknown end users. Various organizations including research organizations, job fair organizers, religious organizations, also send spam to unknown users. The organizers are getting thousands of end user's mail accounts from other organizers to spam end users. Mostly organizers don't provide the facility to unsubscribe from their mailing list.

Regardless of end users contacts and activities, email accounts created several years ago receive more spam than others. To analyze this, we have selected 20 email accounts from our network and monitored the spam traffic. Out of 20, ten email accounts are four years old, in which five accounts are heavy users accounts and remaining five users are non heavy users accounts. The heavy users maintain own website, blogs and actively involve in various forums and chat rooms [6]. The selected heavy users email accounts were published in various newspaper advertisements and others. Remaining 10 email accounts are new email accounts created 14 months ago. In this category, 5 mail accounts belong to heavy users, the remaining 5 accounts are non heavy users. We have monitored the spam traffic received by these 20 accounts for a period of 6 weeks from Jan 1, 2007 to Feb 15, 2007. The spam traffic is illustrated in the table.3**.**

When we say old email account it means that accounts were created four years ago. From our study, we had found out that old heavy users email accounts attracted more spam than others. The data collection for six weeks is illustrated in figure 4. The x axis represents weeks from January 1 to February 15. The Y axis is the number of spam received by end users. The four years old heavy users email account received spam ranging from 78 to 1015, which is an average of 532 per week. The four year old heavy users email accounts attracted roughly 45% more spam than four years old light users mail accounts. Since the network suffered by DDoS attack in the fourth week, it has received more spam.

**Table 3**: Number of spam received by all types of email accounts

| User | Number of Spam received | | | | | |
|---|---|---|---|---|---|---|
| | Week 1 | Week 2 | Week 3 | Week 4 | Week 5 | Week 6 |
| Old Heavy user 1 | 78 | 756 | 408 | 1025 | 98 | 665 |
| Old Heavy user 2 | 299 | 410 | 544 | 877 | 310 | 590 |
| Old Heavy user 3 | 252 | 1015 | 862 | 1045 | 142 | 890 |
| Old Heavy user 4 | 157 | 864 | 589 | 920 | 189 | 601 |
| Old Heavy user 5 | 283 | 929 | 544 | 1200 | 320 | 989 |
| Average | 213.8 | 794.8 | 589 | 1013 | 211 | 747 |
| Old Light User 1 | 456 | 302 | 952 | 1089 | 478 | 319 |
| Old Light User 2 | 377 | 194 | 589 | 640 | 345 | 225 |
| Old Light User 3 | 550 | 151 | 136 | 768 | 490 | 187 |
| Old Light User 4 | 141 | 863 | 589 | 987 | 167 | 641 |
| Old Light User 5 | 141 | 108 | 45 | 234 | 128 | 120 |
| Average | 333 | 323 | 462 | 892 | 321 | 298 |
| New Heavy user 1 | 0 | 0 | 0 | 15 | 0 | 0 |
| New Heavy user 2 | 0 | 0 | 0 | 8 | 0 | 0 |
| New Heavy user 3 | 0 | 0 | 0 | 10 | 0 | 0 |
| New Heavy user 4 | 0 | 0 | 0 | 14 | 0 | 0 |
| New Heavy user 5 | 0 | 0 | 0 | 12 | 0 | 0 |
| Average | 0 | 0 | 0 | 11 | 0 | 0 |
| New Light User 1 | 0 | 0 | 0 | 9 | 0 | 0 |
| New Light User 2 | | | | 8 | 0 | 0 |
| New Light User 3 | 0 | 0 | 0 | 9 | 0 | 0 |
| New Light User 4 | 0 | 0 | 0 | 7 | 0 | 0 |
| New Light User 5 | 0 | 0 | 0 | 2 | 0 | 0 |
| Average | 0 | 0 | 0 | 7 | 0 | 0 |

The four year old light users (non heavy users) mail id received more spam similar to heavy user's id, ranges from 45 to 952 averages 372. The 4 years old non heavy user's mail ids attracted less spam than 4 years old heavy users' id as shown in figure. Regardless of users characters whether heavy or light user, the old email ids attracted more spam than 14 months old new email accounts. Figure 5, shows that, the 14 months old email account did not receive spam. In figure, the x axis is week and y axis is the number of spam received by end users account. Roughly the spam ranges from 0 to 11 with an average rate of 2.

The 14 months old relatively new accounts didn't attract spam except DDoS attack period. Since the network suffered by DDoS attack in fourth week, it has received spam mails as shown in figure. The 14 months old heavy users email accounts didn't receive spam similar to 14 months old light user's mail accounts.

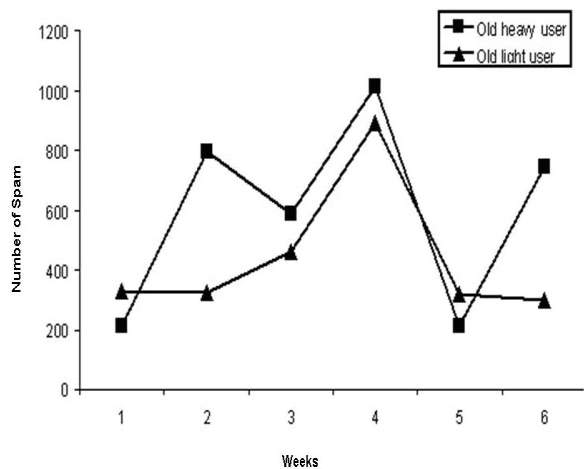

**Figure.4**. Average number of spam received by old heavy users and old non heavy users accounts for 6 weeks

Regardless of heavy user or light user, the 14 month old new mails didn't receive spam. Literally the new email accounts are free from spam for particular period of time, in our case it is up to 14 months. 14 months old heavy users' id received more spam than 14 months old light users' accounts during DDoS attack as shown in figure.5.

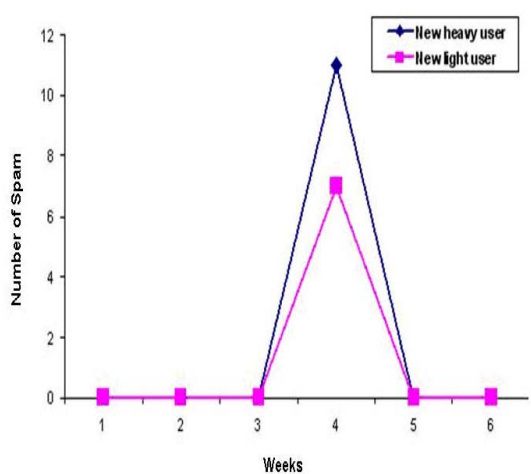

**Figure.5**.Average number of spam received by new heavy users and new non heavy users accounts for 6 weeks from January 1 to February 15

## 6. Conclusions

From our study we conclude that spam can be classified into 2 wide categories. The first category is spam with attachment and the second category is spam without attachments. Spam without attachment are text messages and links to the intended target. Spam with attachments can be classified into 4 types such as Spam mails containing image file (.gif) and text message, spam containing image, text, URL, spam containing only image with clickable web link and spam containing worms, virus, and trojans as an attachments. Spam without attachments are small in size but spam with attachment is bigger in size. The volume of spam traffic is not related to legitimate mail traffic. We have also analyzed the types of email accounts that attract more spam. We conclude from our study that old heavy user's email accounts receive more spam than relatively new email accounts. The four years old heavy users email accounts received 45% more spam than four years old light users accounts. The 14 months old relatively new accounts didn't attract spam except DDoS attack period.

### Acknowledgement

This work was supported by the second stage of the Brain Korea 21 Project in 2007.